# SequenceLab: A Comprehensive Benchmark of Computational Methods for Comparing Genomic Sequences

**Maximilian-David Rumpf[1,#], Mohammed Alser[2,#,*], Arvid E. Gollwitzer[1,*]**

**Joël Lindegger[1], Nour Almadhoun[1], Can Firtina[1], Serghei Mangul[3], Onur Mutlu[1,*]**

[1]Department of Information Technology and Electrical Engineering, ETH Zürich, 8092 Zurich, Switzerland

[2]Department of Computer Science, Georgia State University, Atlanta, GA, USA

[3]Department of Clinical Pharmacy, University of Southern California, Los Angeles, CA, 90089, USA

[#]These authors contributed equally to this work.

*Corresponding author. Department of Information Technology and Electrical Engineering, ETH Zurich, Gloriastrasse 35, 8092 Zurich, Switzerland.

E-mail: arvidg@ethz.ch (A. E. G.), malser@gsu.edu (M.A.), omutlu@ethz.ch (O. M.)

## Abstract

Computational complexity is a key limitation of genomic analyses. Thus, over the last 30 years, researchers have proposed numerous fast heuristic methods that provide computational relief. Comparing genomic sequences is one of the most fundamental computational steps in most genomic analyses. Due to its high computational complexity, optimized exact and heuristic algorithms are still being developed. We find that these methods are highly sensitive to the underlying data, its quality, and various hyperparameters. Despite their wide use, no in-depth analysis has been performed, potentially falsely discarding genetic sequences from further analysis and unnecessarily inflating computational costs. We provide the first analysis and benchmark of this heterogeneity. We deliver an actionable overview of the 11 most widely used state-of-the-art methods for comparing genomic sequences. We also inform readers about their advantages and downsides using thorough experimental evaluation and different real datasets from all major manufacturers (i.e., Illumina, ONT, and PacBio). *SequenceLab* is publicly available at https://github.com/CMU-SAFARI/SequenceLab.

## Introduction

The rapid advancement of genomics and sequencing technologies continuously calls for adjusting existing algorithmic techniques or developing completely new computational methods across diverse biomedical domains[1]. Omics algorithms implemented as computational tools help life-science and biomedical researchers analyze increasingly complex data, solve complex biological problems, and lay the groundwork for novel clinical translations. One of the key success factors of computational omics software tools is how fast an omics tool can solve the subject problem while accounting for some technical aspects of the biotechnological protocol[2,3]. As we witness an explosive growth of omics data generated in laboratories worldwide, high computational complexity and limited computing infrastructure remain fundamental constraints to scaling analyses[3–6].

One of the most fundamental computational steps in most genomic analyses is comparing genomic sequences to find their differences and similarities. Comparing genomic sequences is a key step in, for example, DNA and RNA sequence alignment[7–16], taxonomic profiling of metagenomics samples[17–19], viral quasispecies[20], bisulfite-converted sequences[21–23], reconstructing ancient genomes[24], B and T-cell receptor repertoire analysis[25,26], and building population-specific reference genomes[27,28]. Comparing genomic sequences has long been a major bottleneck and a computational challenge for two reasons: First, the analyzed datasets are typically large. Modern sequencing machines generate genomic sequences at an exponentially higher rate than prior sequencing technologies, with their growth far outpacing the growth in computational power in recent years[3–5]. For example, the Illumina NovaSeq 6000 system can generate about 136 gigabases per hour per instrument[29–35].
In contrast, the PromethION 48 system from Oxford Nanopore Technologies (ONT) generates 1.4x higher throughput than that, and its throughput is constantly improving[36]. Second, calculating the optimal number of differences between two genomic sequences requires using dynamic programming (DP) algorithms[15,37] such as Levenshtein distance[38], Smith-Waterman[39], and Needleman-Wunsch[40]. Such DP-based algorithms are well-known for having quadratic time and space complexity in the genomic sequence length, which is inefficient for processing large genomic data. To detect all edits resulting from genomic alteration[41,42] or sequencing errors[43,44], quadratic time (DP-based) approaches are unavoidable, as optimal genomic comparison is impossible in strong sub-quadratic complexity[45,46]. To reduce the workload of such quadratic time algorithms, many heuristic methods are used to quickly quantify the similarity of a pair of genomic sequences, mainly known as *filtering* techniques[47–52], to reduce the workload of such quadratic time algorithms. Most of these techniques make binary decisions to accept or reject the genomic sequence pair before using the costly DP algorithms between these sequences.[1,53–70]

Over the past several decades, many exact algorithms and heuristics have been proposed to accelerate comparing genomic sequences. A heuristic is a practical method that is not optimal in all cases but aims to provide close-to-optimal solutions at greater speed[71]. Though heuristics usually reduce the execution time and memory footprint of an application, heuristics can compromise accuracy, limit the reproducibility and usability of the adopted bioinformatics methods, and quickly become obsolete due to changes in the properties of the underlying genomic data. However, these pitfalls are not well characterized – preventing an educated choice of adequate heuristics for a given application. Our goal is to provide the first comprehensive benchmark of a prominent set of heuristic algorithms used for quickly comparing genomic sequences such that it guides a wide range of users in academia and industry toward selecting the heuristic that suits their needs.[13,72–101]

We make five key contributions to achieve this: 1) We provide the *first* comprehensive analysis of algorithmic foundations and methodologies of the 11 most widely used algorithms in read mapping for taxonomic profiling of metagenomic samples. 2) We provide a rigorous experimental evaluation to examine the algorithms' speed and accuracy. Heuristic methods for comparing genomic sequences often greatly relieve the computationally expensive subsequent steps but can come at the cost of removing sequences from downstream analysis and excessively large filtering time. In this context, we consider the heuristic methods for comparing genomic sequences as pre-alignment filters. 3) We determine whether a combination of heuristic methods can perform better than a single method alone. 4) We analyze empirically and theoretically how the properties (e.g., number, distribution, and type of genomic differences) of the genomic data and sequence length affect filtering performance and accuracy. We use real datasets representing the three prominent sequencing technologies, Illumina, PacBio HiFi, and ONT, with a wide range of edit distance thresholds to control the filtering rigor. This analysis is tailored towards providing bioinformatics researchers a firm understanding of which filtering heuristic offers the best performance for a given real-world application. 5) Our study provides a plug-and-play benchmark suite for evaluating the performance of existing and future algorithms. Enabling rapid benchmarking can vastly accelerate further research and add to the already plentiful applications of filters. It will also identify the techniques or compare genomic sequences that balance accuracy and speed well. We hope that it catalyzes new research in genome sequence analysis and bioinformatics.

## Methods

We survey the existing algorithmic methods used for comparing genomic sequences. We provide a comprehensive overview of computational heuristic methods for comparing genomic sequences published from 1993 until 2020 (**Table 1**). To assess them and evaluate their pros and cons, we develop the *first* benchmark suite, *SequenceLab*, that includes implementations of state-of-the-art algorithmic methods. We observe that there is a critical need for such a benchmark suite due to five main reasons: 1) Without having an available benchmark suite, it is challenging to understand which algorithm is best to be utilized in omics tools based on, for example, developer needs, target platform, and accuracy. 2) Each algorithm is implemented using a different programming language, which leads to differences in performance and difficulties in integrating such implementations with existing projects. 3) Some algorithms are implemented as part of large tools, and it is time-consuming to understand the code of the complete tool to be able to isolate the algorithm implementation. 4) Some algorithms are implemented using hardware architectures; running them requires owning the target hardware platform (e.g., expensive FPGA or GPU) or implementing the algorithm as a CPU program from scratch. 5) Some algorithms do not have an available implementation as they are closed source, the link to the source code is invalid, or their published manuscript does not include a link to the source code.

To address these five challenges, we build SequenceLab, which includes implementations of 11 state-of-the-art algorithmic methods for comparing genomic sequences. The filter benchmark suite is written in C to be compatible with most bioinformatics tools written in C and C++. It accepts input data that contains sequence pairs in a versatile format that can be generated from, e.g., a minimap2's PAF file. SequenceLab provides four different types of output data. 1) The raw total execution time for each tool. 2) a table where each index of its rows corresponds to the index of its corresponding input sequence pair, and the columns represent the outputs of the included algorithmic method as 1 or 0. A cell with a value of 1 means that the tool accepts the sequence pair given the estimated edit distance is less than or equal to the user-defined edit distance threshold. A cell with a value of 0 means that the tool rejects the sequence pair. Heuristic algorithms for comparing genomic sequences can quickly estimate the edit distance between two sequences. Prior work has shown that exact edit distance calculation is computationally expensive, and it is impossible to develop an edit distance algorithm with strong sub-quadratic time[45]. If the estimated edit distance exceeds a user-defined threshold, the input sequence pair is considered dissimilar and excluded from further analysis. In genomic studies, only sequences with an edit distance less than or equal to a user-defined threshold $E$ are considered biologically useful[1,47,102]. The user-defined edit distance threshold should be high enough to accommodate possible sequencing errors (with a rate of 0.1–20 % of the read length, depending on the sequencing technology) and genetic variations specific to the individual organism's DNA[13]. 3) Statistics on workload overlap, which indicates how many methods agree on the decision of acceptance/rejection. This helps inform developers on which algorithmic method can be executed after which algorithmic method for a more robust multi-level filtering mechanism.

4) Evaluation statistics such as False Accept Rate (FAR) and False Reject Rate (FRR). Developing an ideal comparison algorithm requires achieving an FRR of 0 and a minimal FAR. We define the FAR and FRR as follows:

$$False\ Accept\ Rate\ (FAR) = \frac{Number\ of\ False\ Accepts}{Number\ of\ False\ Accepts + Number\ of\ True\ \ Rejects}$$

$$False\ Reject\ Rate\ (FRR) = \frac{Number\ of\ False\ Rejects}{Number\ of\ False\ Rejects + Number\ of\ True\ \ Accepts}$$

Rejecting some correct sequence pairs, i.e., FRR > 0, is generally undesirable as a correct pair is excluded from downstream analyses. For read mapping, this could mean that some genetic variants can be missed.

### There Are Four Main Approaches to Genomic Sequence Comparison

There are four main approaches to cluster existing methods: base counting, q-gram, Pigeonhole Principle, and sparse dynamic programming (Sparse DP). Researchers have proposed filters for CPU (most common), GPUs[103], FPGAs[104,105], and Processing-in-Memory systems[47]. Each accelerator has tradeoffs, but their evaluation is outside the scope of this paper.

**Table 1: Overview of the methods surveyed**. Not all are included in the benchmarking suite due to non-functional or missing source code. Some filters can only be used as part of a larger pipeline and are inaccessible for standalone use cases.

| Name | Year | Approach | Short/ Long Reads | Native Platform | Language | URL Software |
|---|---|---|---|---|---|---|
| QCKer[106,107] | 2020 | q-gram | Short | CPU/ FPGA | C | N/A |
| SneakySnake[105] | 2019 | Pigeonhole | Short/ Long | CPU/ GPU/ FPGA | C/ C++ | https://github.com/CMU-SAFARI/SneakySnake |
| Shouji[102] | 2019 | Pigeonhole | Short | FPGA | C/ Verilog | https://github.com/CMU-SAFARI/Shouji |
| GenCache III: Banded Myers[108] | 2019 | Sparse DP | Short | CPU | | N/A |
| GenCache II: SHD[108] | 2019 | Pigeonhole | Short | CPU | | N/A |
| GenCache I: HD[108] | 2019 | Pigeonhole | Short | CPU | | N/A |
| Chaining in minimap2[8] | 2018 | Sparse DP | Short/ Long | CPU | C/ Python | N/A |
| GRIM-Filter[47] | 2018 | q-gram | Short | PIM | C | https://github.com/CMU-SAFARI/GRIM |
| Bitmapper2[109] | 2018 | q-gram | Short | GPU | C | N/A |
| D-Soft (Darwin)[48] | 2018 | q-gram | Long | FPGA | C++ | https://github.com/yatisht/darwin |
| MAGNET[49] | 2017 | Pigeonhole | Short | CPU | Matlab | https://github.com/BilkentCompGen/MAGNET |
| GateKeeper[104] | 2017 | Pigeonhole | Short | FPGA | C/ Verilog | https://github.com/BilkentCompGen/GateKeeper |
| PUNAS (improved SHD)[110] | 2017 | Pigeonhole | Short | CPU (AVX2/KNL) | | https://github.com/Xu-Kai/PUNASfilter |
| rHAT[111] | 2016 | q-gram | Long | CPU | C++ | https://github.com/HIT-Bioinformatics/rHAT |
| SHD[103] | 2015 | Pigeonhole | Short | SIMD | C/ SIMD | https://github.com/CMU-SAFARI/Shifted-Hamming-Distance |
| Bitmapper[112] | 2015 | Pigeonhole | Short | CPU | C | N/A |
| mrsFAST-Ultra II[113] | 2014 | Pigeonhole | Short | CPU | C | http://mrsfast.sourceforge.net |
| mrsFAST-Ultra I[113] | 2014 | Base Counting | Short | CPU | C | http://mrsfast.sourceforge.net |
| RazerS 3 I: SWIFT[114] | 2012 | q-gram | Short | CPU | C++ | http://www.seqan.de/projects/razers |
| RazerS 3 II: Pigeonhole[114] | 2012 | Pigeonhole | Short | CPU | C++ | http://www.seqan.de/projects/razers |
| Hobbes[115] | 2011 | q-gram | Short | CPU | C++ | http://hobbes.ics.uci.edu |
| SHRiMP 2[116] | 2011 | q-gram | Short | CPU | C | http://compbio.cs.toronto.edu/shrimp |
| GASSST II: Base Counting[117] | 2010 | Base Counting | Short | CPU | C++ | https://www.irisa.fr/symbiose/projects/gassst |
| GASSST I: Tiled-NW[117] | 2010 | Sparse DP | Short | CPU | C++ | https://www.irisa.fr/symbiose/projects/gassst |
| Adjacency Filtering (mrFAST)[51] | 2010 | Pigeonhole | Short | CPU | C | https://github.com/BilkentCompGen/mrfast |
| ZOOM[118] | 2008 | q-gram | Short | CPU | | https://www.bioinfor.com/zoom-1-3-gui-release-next-gen-seq |
| GenCache[119] | 2005 | q-gram | Short | CPU | C++ | https://bibiserv.cebitec.uni-bielefeld.de/swift |
| Better Filtering with Gapped q-Grams[120] | 2001 | q-gram | Short | CPU | | N/A |
| FLASH[121] | 1993 | q-gram (LUT) | Short | CPU | C | N/A |

### Base Counting Is the Simplest Method for Comparing Genomic Sequences

Base counting compares the frequency of individual genomic bases between two sequences. For DNA, this means counting the occurrence of adenine (A), thymine (T), guanine (G), and cytosine (C) bases in each sequence. By calculating the Manhattan distance between the respective base counts, we can derive a lower bound to the edit distance of these two sequences (**Figure 1**).

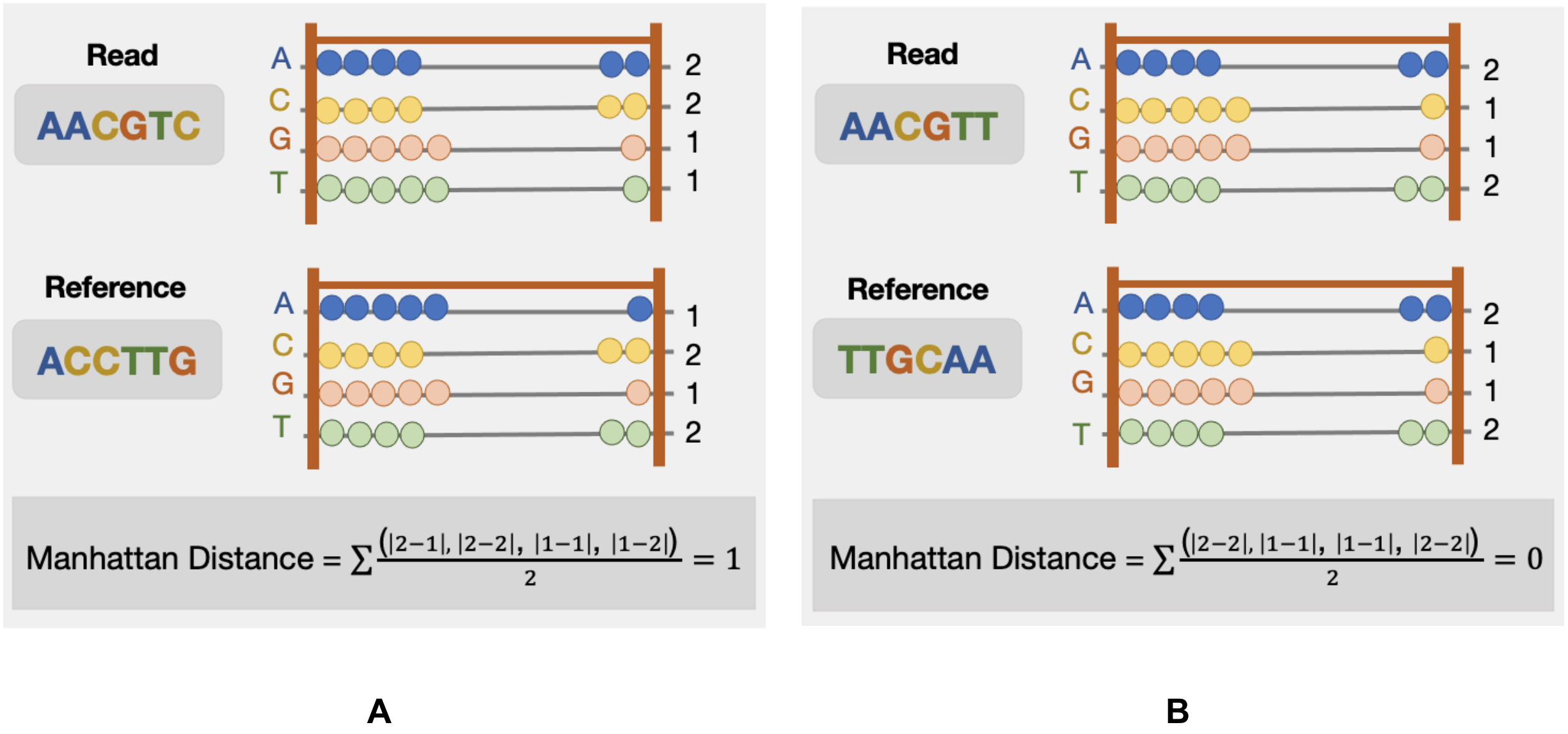

**Figure 1: Base Counting approach for comparing genomic sequences.**

A) It counts every two differences as, at most, one edit. B) No edit is detected despite the six differences between the two sequences, as both sequences have the same number of As, Cs, Gs, and Ts.

The base counting algorithm does not provide an overestimation of the number of differences between two sequences for two reasons. 1) The Manhattan distance divides the sum of the total number of differences between the counts by two as it assumes an increase in one of the counts for As, Cs, Gs, or Ts leads to a decrease in one of the other counts (given that the lengths of the two sequences are equal). The increase and the decrease in counts can be caused by a single change (e.g., deleting one character) or two changes (e.g., two substitutions). Hence, the base counting algorithm counts them conservatively as, at most, a single difference (**Figure 1A**). 2) It does not consider the location of the base in the sequence when counting the bases. This means that swapping two characters in one of the sequences is not counted as two differences. For example, the base counting algorithm considers "AACGTT" the same as "TTGCAA" (**Figure 1B**). Hence, the base counting algorithm provides a zero False Reject Rate (FRR).

The computational complexity of base counting is $O(n)$ with $O(1)$ space, where $n$ is the sequence length. There is a key potential for accelerating the different implementations of the base counting algorithm as the algorithm is simple and highly parallelizable (counting one base is independent of counting the other bases). Using parallelization, its computational complexity drops to $O(log(n))$ with $O(log(n))$ space. If we query a sequence repeatedly, building an index for constant lookups may provide further speedup. It is important to note that the complexity is independent of the edit distance threshold chosen, which is advantageous for long sequences.

**Q-gram Improves Sequence Comparison Accuracy by Counting Subsequences Instead of Bases**

The q-gram method can be seen as an extension of the base counting approach. Q-gram methods compare the abundance of q-long subsequences to calculate a lower-bound estimate of the edit distance. Much like Base Counting, these approaches are designed to have a zero False Reject Rate. The old implementations[122] of q-gram approaches enumerate all subsequences of a certain length, $q$, and quantify the occurrence of each of such enumerations. There are four possible enumerations for $q$=1, 16 enumerations for $q$=2, 64 enumerations for q=3, 256 enumerations for $q$=4, and 1024 enumerations for $q$ =5. For long q-grams, this approach becomes infeasible; $4^{15} \approx 10^{9}$ enumerations for $q$=15. Recent implementations extract the q-gram from the sequence itself instead of enumerating all possible subsequences. For a sequence of length $n$, there are $n - q + 1$ overlapping q-grams. An edit operation in any position can modify a maximum of $q$ q-grams. Thus, if two sequences have fewer than $n - q + 1 - q * E$ matching q-grams, they are guaranteed to have more than $E$ edits (**Figure 2**).

Q-grams (or k-mers) still play a large role in many biological algorithms, such as seeding or Metagenomics[8,17,123]. They are straightforward to implement in their simplest form. One major advantage of the q-gram approach is that its complexity does not depend on the edit distance threshold. Thus, its computational complexity is similar to that of the base counting approach $O(n)$ (without using any data structure) for a sequence length of $n$. Q-gram methods are suitable for acceleration due to q-gram independence and the simplicity of the required operations. The key limitation of the q-gram approach is that it does not respect the order in which q-grams appear in the original sequences, which can confuse the decision on whether or not the two compared sequences are similar.

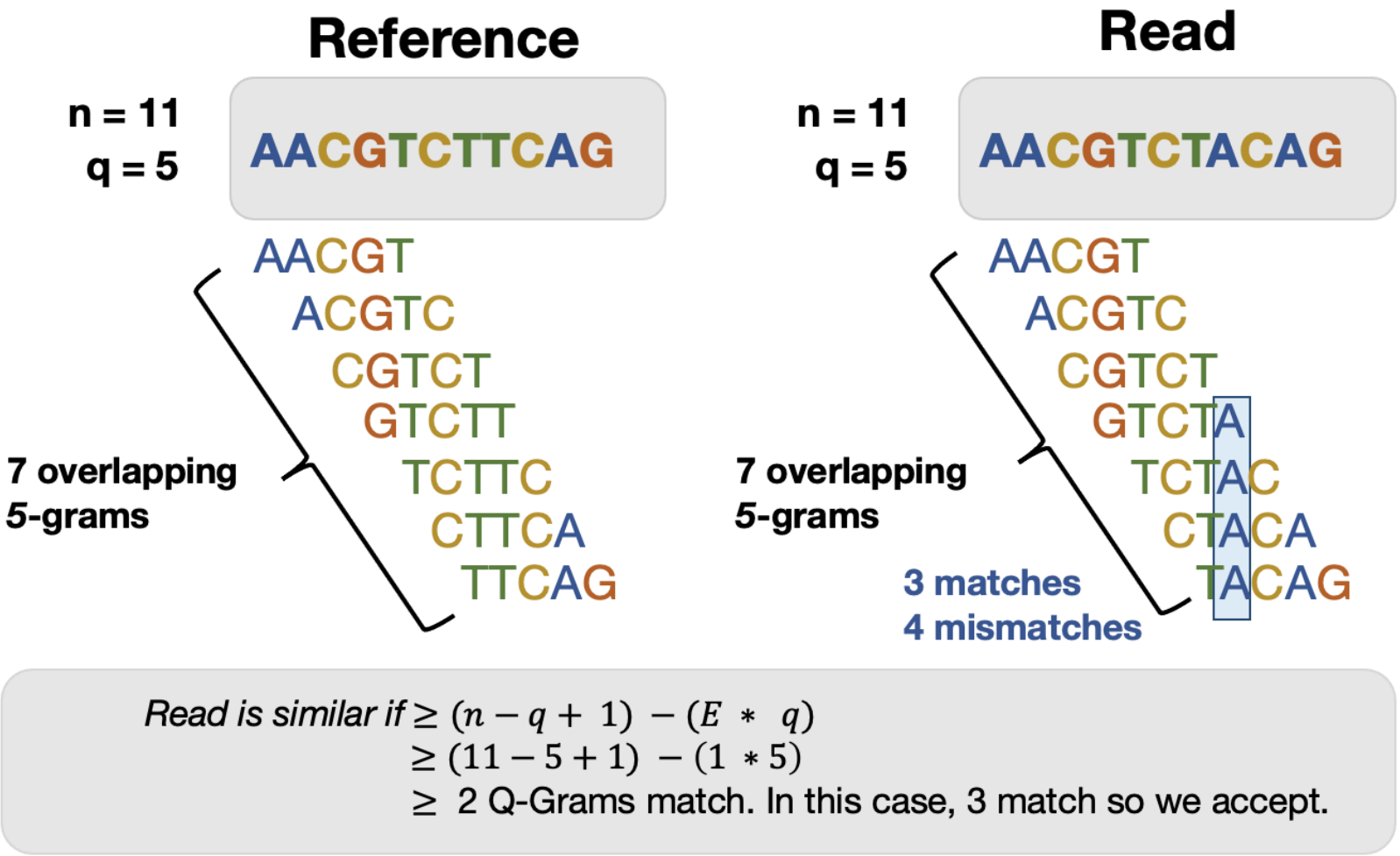

**Figure 2:** q-gram (k=5) filtering example with threshold $E = 1$.

### Pigeonhole Principle Approach Increases Algorithm Complexity

The pigeonhole principle states that if $n$ objects are placed in $k$ boxes with $k < n$, at least one of the boxes contains $ceil(\frac{n}{k})$ objects. Many tools use this principle to determine whether substitutions exist in a sequence. If a substitution exists, there is at least one position where the two sequences differ. To also handle insertions and deletions, many approaches use shifted copies of the sequences[102–105,110].

The simplest version is the Hamming distance algorithm. It determines whether two sequences match by iterating over the sequences and counting the number of times the corresponding characters are different. If there is a single substitution ($n$=1) in one of the sequences, then the single substitution divides the exact matches into two sets ($k$=2) of matching subsequences. Quantifying these matching subsequences can also provide insights into the number of differences based on the Pigeonhole Principle. Unlike Base Counting and q-gram approaches, the Hamming distance approach considers the location of the matching characters in making the similarity decision. Notably, Hamming distance cannot handle insertions and deletions. An insertion operation, for example, is the equivalent of shifting the trailing subsequence of the original sequence to the right. Thus, Hamming distance cannot correctly match the characters in such subsequences. Shifted-sequence approaches address this by also comparing against a right-shifted (for insertion) and left-shifted (for deletions) version of the sequence. Comparing with two right-shifted copies and two left-shifted copies helps to detect at most two operations of the same type (either insertion or deletion) occurring one after another (**Figure 3**). Such a number of shifted copies can also help to detect other cases, such as one insertion followed (not necessarily consecutively) by a deletion followed by an insertion. These three edit operations have $E$+1=4 matching subsequences distributed among the shifted copies.

The computational complexity of the Pigeonhole Principle varies based on the implementation. Except for Hamming distance, the Pigeonhole Principle approach heavily relies on the edit distance threshold $E$, a conservative parameter used to control the maximum number of operations of the same type that occurs after each other and can be detected by the Pigeonhole Principle approach. Most methods have a complexity of $O(nE)$ where $n$ is the sequence length. The value $E$ usually depends on the type of sequencing data used for the analysis. For example, $E$ of 5% of read length suffices for Illumina as Illumina data usually has very high sequencing accuracy that obviates the need to accommodate them via higher $E$ value[13,124].

Algorithms such as Hamming distance can be simple and only use bitwise comparison operations (i.e., XOR circuit), which makes it attractive for acceleration. More complex algorithms that generate shifted copies of a sequence can still be attractive for acceleration as they rely on shift operation and bitwise comparisons and have been widely implemented in GPUs, FPGAs, and Processing-in-Memory.

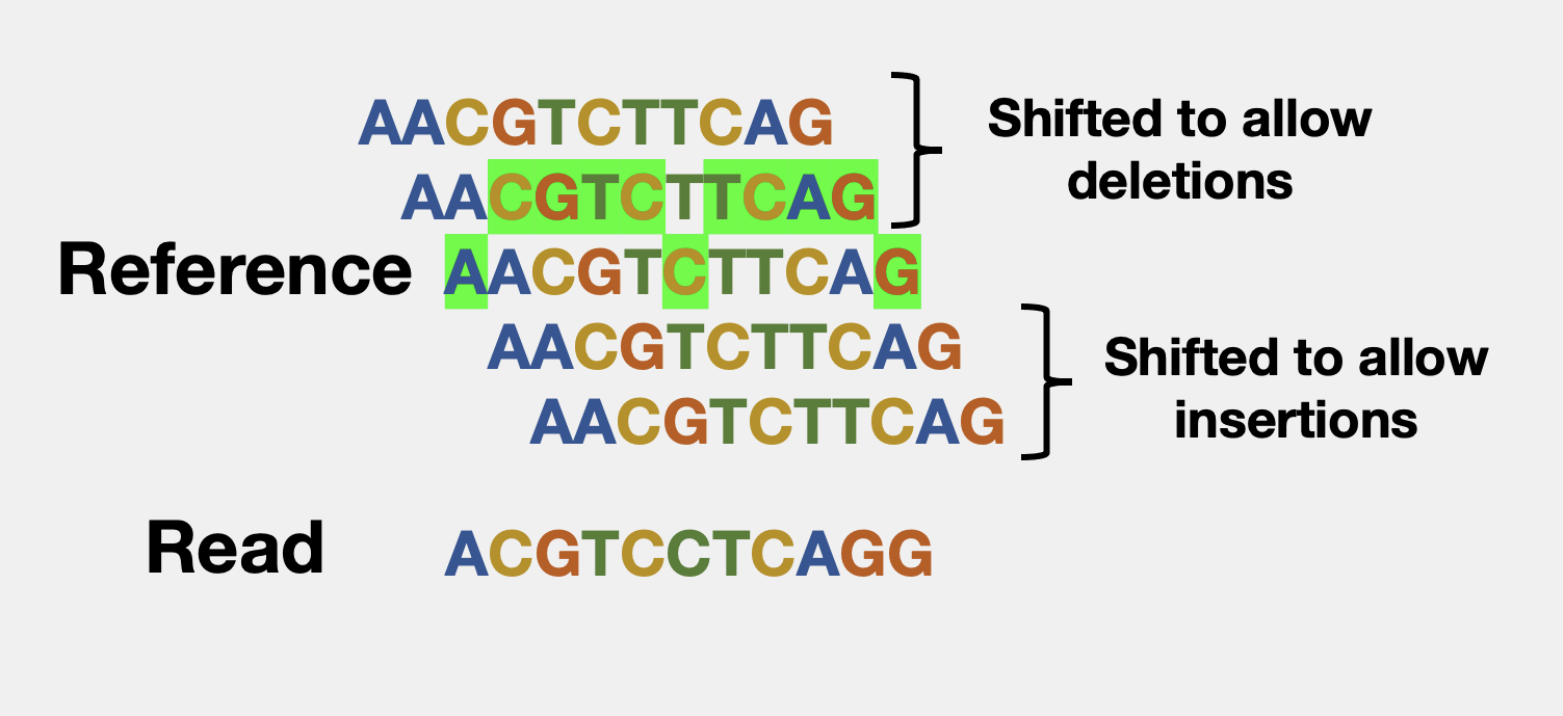

**Figure 3:** Pigeonhole Principle-based filter that uses shifted sequences to tolerate up to two insertions and deletions. The highlighted sections mark matches with the read sequence. We see that it tolerates the deletion at position two.

**Sparse Dynamic Programming Uses Expensive Alignment Sparingly**

Sparse Dynamic Programming (DP) algorithms utilize the concept of exact matches, or seeds, between two sequences to optimize execution time. The algorithms find exactly matching subsequences between two sequences and only use expensive edit distance calculations for the gaps between every two matching subsequences (**Figure 4**). If these gaps are sufficiently small, overall runtime is greatly improved compared to expensive edit distance calculations on the entire sequence. This approach is very useful in popular software such as minimap2[8] and rHAT[111].

Recent advancements in the field have led to the development of GPU and FPGA accelerators for the chaining algorithm, resulting in 7x and 28x acceleration, respectively, compared to the sequential implementation executed with 14 CPU threads in minimap2[8]. However, the worst-case time and space complexity for this approach is still $O(n^2)$ where $n$ is the length of the sequence.

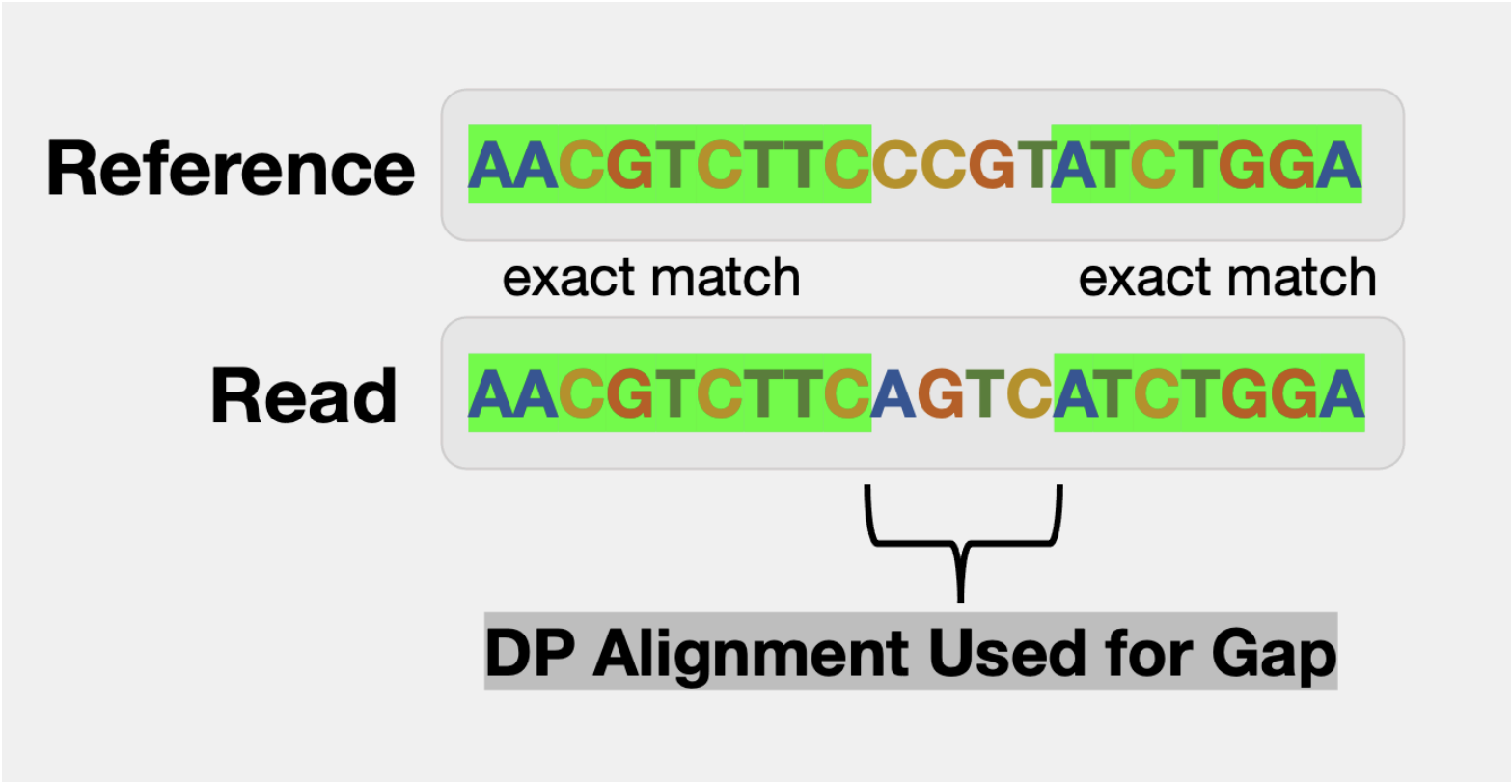

**Figure 4:** Sparse Dynamic Programming. Expensive alignment is only performed between the exact matching sections.

## Evaluation Results

We evaluate every algorithm included in SequenceLab using six key performance metrics: specificity, false accept rate (FAR), false reject rate (FRR), filtering throughput, execution time, and workload overlap. We define throughput as the number of sequence pairs examined per second; thus, a higher throughput is better. We define specificity as 1 - FAR; thus, a higher specificity is better. We also compare the performance of these filters with that of two state-of-the-art sequence aligners, Edlib[125] and KSW2[8,126], whenever applicable. We evaluate performance using different real datasets representing the three prominent sequencing technologies: Illumina, PacBio HiFi, and ONT. The filtering accuracy depends greatly on the edit distance threshold and the data analyzed. If a filter has zero FRR and we apply a threshold of 5%, it is guaranteed to accept all pairs with an edit distance of less than (or equal) 5% (as a percentage of read length). We test the thresholds from 0% to 10% for both Illumina and HiFi data and from 0% to 20% for ONT data. This covers practical and real-world ranges of edit distance threshold. While higher thresholds would be technically possible, they are rarely biologically useful and thus not seen in real-world applications[1,47,102]. It is important to note that 0% is a special case in this analysis. An edit distance of 0% means that the two sequences must be identical to pass, which can be tested perfectly in $O(n)$ with the Hamming Distance algorithm. Some filters use this special case and run Hamming Distance if threshold 0 is given. Unfortunately, many filters do not and thus unnecessarily exhibit non-zero errors in this special case.

The system configuration we used was a 2.3 GHz Intel Xeon Gold 5118 CPU with 24 cores and 48 threads, along with 192 GB of RAM. To maintain comparability, only CPU implementations were considered in our experiments. Most filters rely on numerous parameters: q-gram length, early termination, sliding window sizes, or algorithm choice. To assess performance fairly, we used the author's recommended parameters. These parameters can be viewed and tweaked within the benchmarking suite we provide.

## Datasets

To provide the most comprehensive overview possible, we examined real sequence datasets from Illumina (SRR10035390, accessed July 2022), PacBio HiFi (SRR12519035, accessed July 2022), and ONT (SRR12564436, accessed July 2022). For each of these datasets, we used minimap2 (version 2.23-r1111) to generate two PAF files – one after chaining and one after mapping, for a total of six datasets. The PAF files generated before mapping (after chaining) contain sequence pairs that still have a high number of edits since filtering is not executed. In contrast, the PAF files generated after mapping mostly contain sequence pairs with a number of edits less than or equal to the edit distance threshold. We chose Homo-Sapiens CHM13 Telomere-to-Telomere[127] (GCA_009914755, accessed July 2022) as our reference genome. We extracted the sequence pairs from these files and further split them into the shortest 90% and longest 10% of sequences to better observe the effect of ultra-long sequences on the performance of the different comparison methods. We extensively analyzed the datasets to understand edit distance distribution and confirm sensible edit distance thresholds. In total, we examined twelve datasets. In the spirit of brevity, not all are plotted in the results, but we provide the data and plotting tools in the supplementary materials.

## Overall Performance: Specificity and Throughput

We first evaluate the overall performance of the 11 filters and the two sequence aligners in terms of both specificity and throughput (**Figure 5**). Ideally, we would like to have a filtering algorithm whose specificity and throughput are as close as possible to the top-right corner of the figure. We make four key observations based on **Figure 5**. 1) There is a clear tradeoff between throughput and specificity when using Illumina short reads. SneakySnake and Hamming Distance break this tradeoff, though Hamming Distance has a non-zero FRR that can be unsuitable for some applications, as detailed in Methods. 2) When filtering very long sequences, i.e., PacBio HiFi and ONT, only Base Counting and SneakySnake (HiFi only) maintain a significant throughput. 3) Only SneakySnake, MAGNET, and q-gram (with a k-mer length of 15) provide high specificity for very long reads. In contrast, Adjacency Filtering, SHD, GateKeeper, Shouiji, short q-gram methods (including GRIM), and Base Counting provide very low specificity. 4) For ONT reads, the aligner Edlib outperforms many filters.

We conclude that filter performance varies strongly based on the underlying sequencing technology. SneakySnake is an attractive candidate for Illumina and HiFi reads. There are no filters that perform well on ONT data.

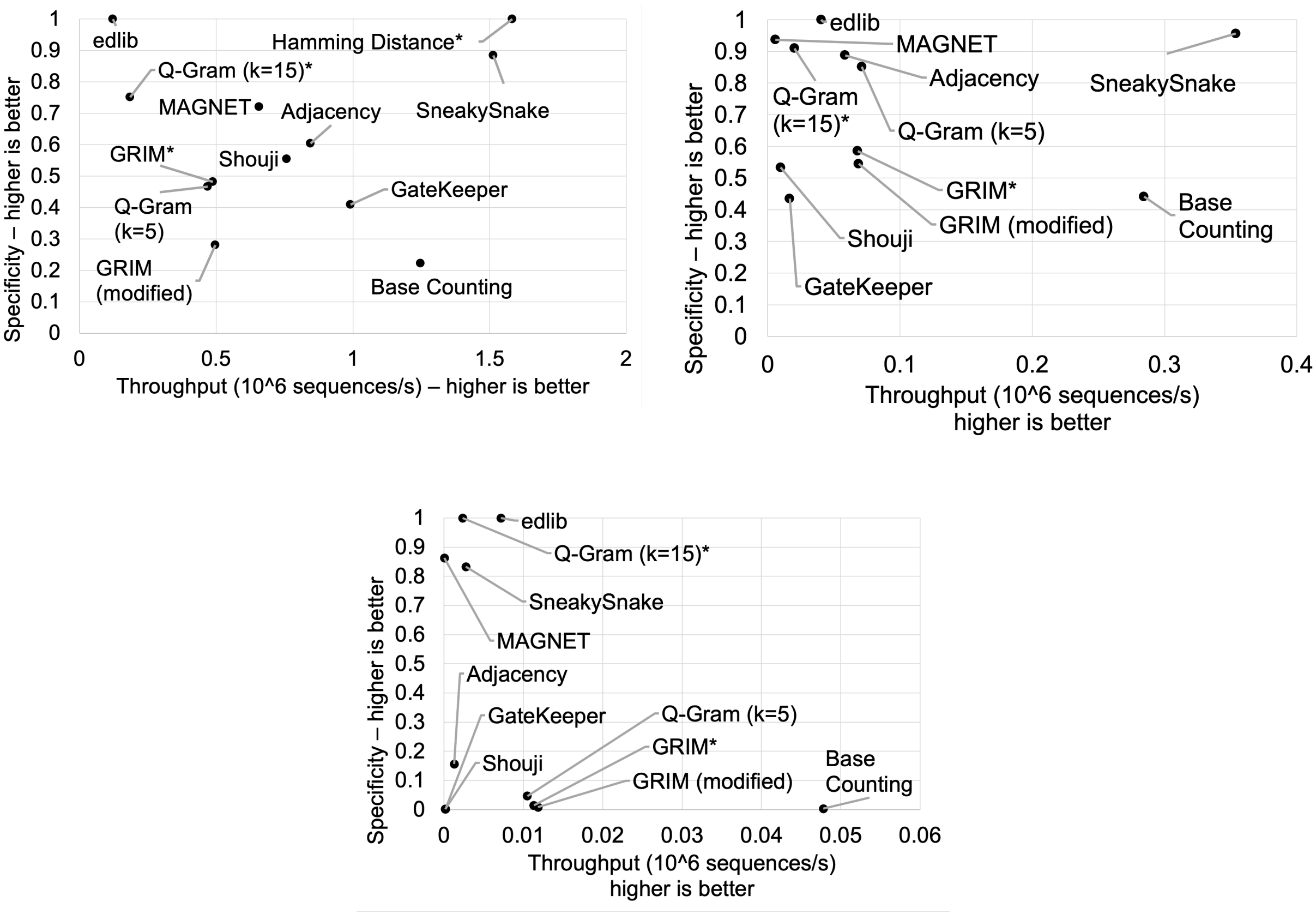

**Figure 5: Scatter plot showing specificity over throughput for Illumina (top left), PacBio HiFi (top right), and ONT (bottom).** The * sign next to the tool name indicates that the filter has False Rejects, which is undesirable and may lead to artificially high specificity. Hamming Distance is excluded from ONT and PacBio HiFi as it removes almost all correct pairs and thus is unsuitable for most applications – see the next section for an in-depth evaluation. SHD was tested but omitted from these charts as it is limited to sequences of length 128 or less.

## False Accept and False Reject Rates

We evaluate the accuracy of the 11 filters using two key metrics: False Accept Rate (FAR) and False Reject Rate (FRR). **Figures 6-8** show the FAR and FRR for six datasets across three sequencing technologies. The left heatmap shows the accuracy for the shortest 90% of sequences and the right for the longest 10% of sequences. Low FAR and FRR are better. A FAR of 1.0 (highest) indicates that the filter does not reject any pairs and accepts everything. An FRR of 1.0 (highest) indicates the filter falsely rejects every correct pair. We make six key observations: 1) FAR sharply increases with higher edit distance thresholds (except Hamming Distance). 2) FAR also increases for longer sequences, most evident in HiFi and ONT datasets. 3) Most filters provide good FAR and FRR for low edit distance thresholds across HiFi and Illumina datasets. SneakySnake provides excellent FAR and zero FRR for all data types up to 10%. 4) Hamming Distance has a potentially acceptable FRR for Illumina reads but rejects 70-100% of correct pairs for HiFi (longest 10%) and ONT data and thus cannot be recommended for general applications. 5) q-gram methods have a very low FAR for HiFi. Q-gram (k=5) is attractive on ONT for edit distance thresholds of 1-5%. 6) SneakySnake and MAGNET are the only filters that maintain low FAR for ONT data (up to a threshold of 8-10%).

We conclude that filtering accuracy highly depends on the edit distance threshold, with many filters becoming ineffective for thresholds above 5%. This observation is vital for application developers, as deploying filters outside this region means no sequences are removed, and total runtime increases. Additionally, we observe that most filters are ineffective at useful edit distance thresholds for long reads.

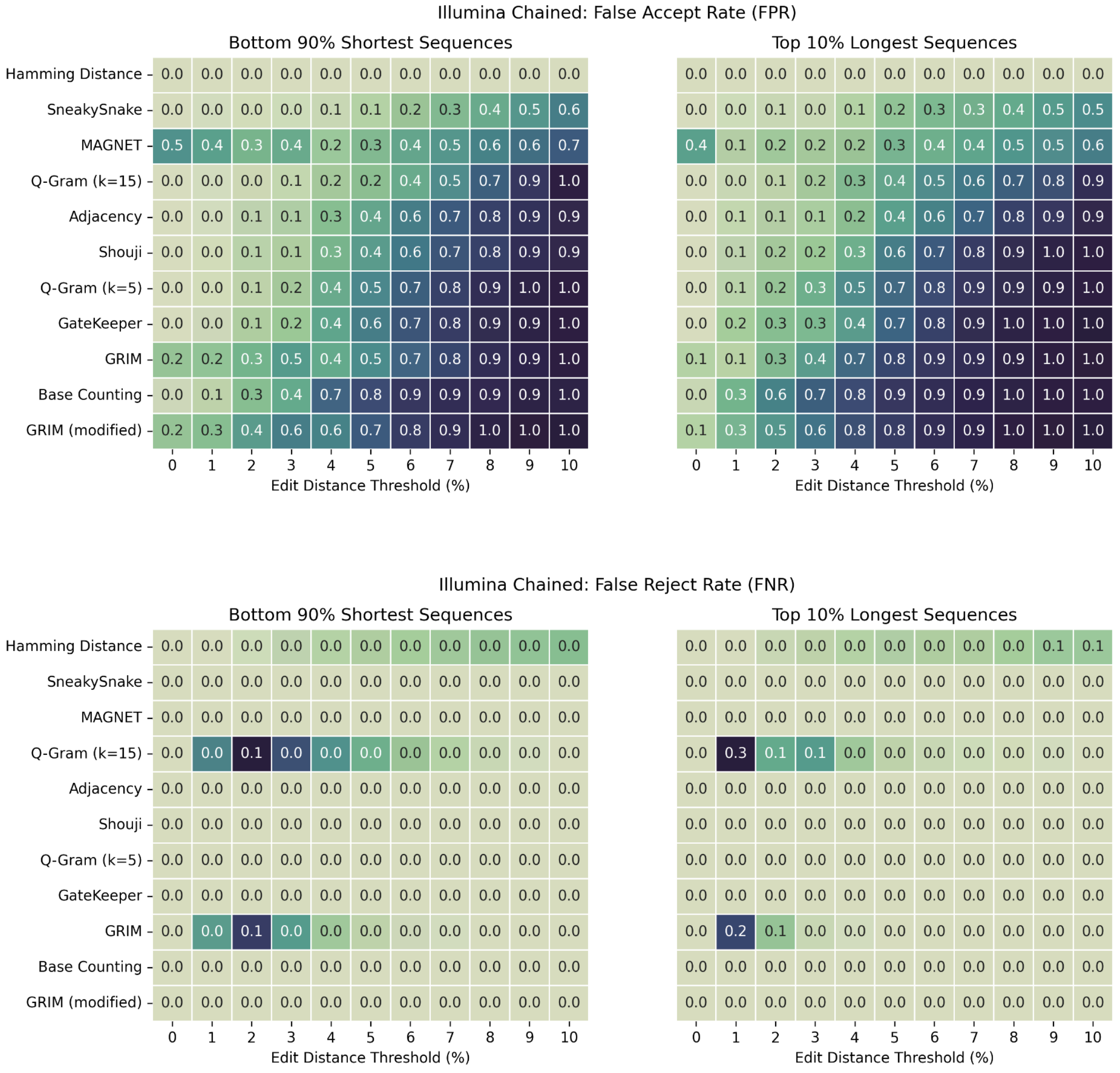

**Figure 6:** False Accept Rate (top) and False Reject Rate (bottom) for Illumina reads after chaining inside minimap2. 0.0 is the best, and 1.0 is the worst score. Note that boosting the False Accept Rate is easy by increasing the False Reject Rate – see Hamming Distance. FRR > 0 is undesirable as some correct pairs may be lost.

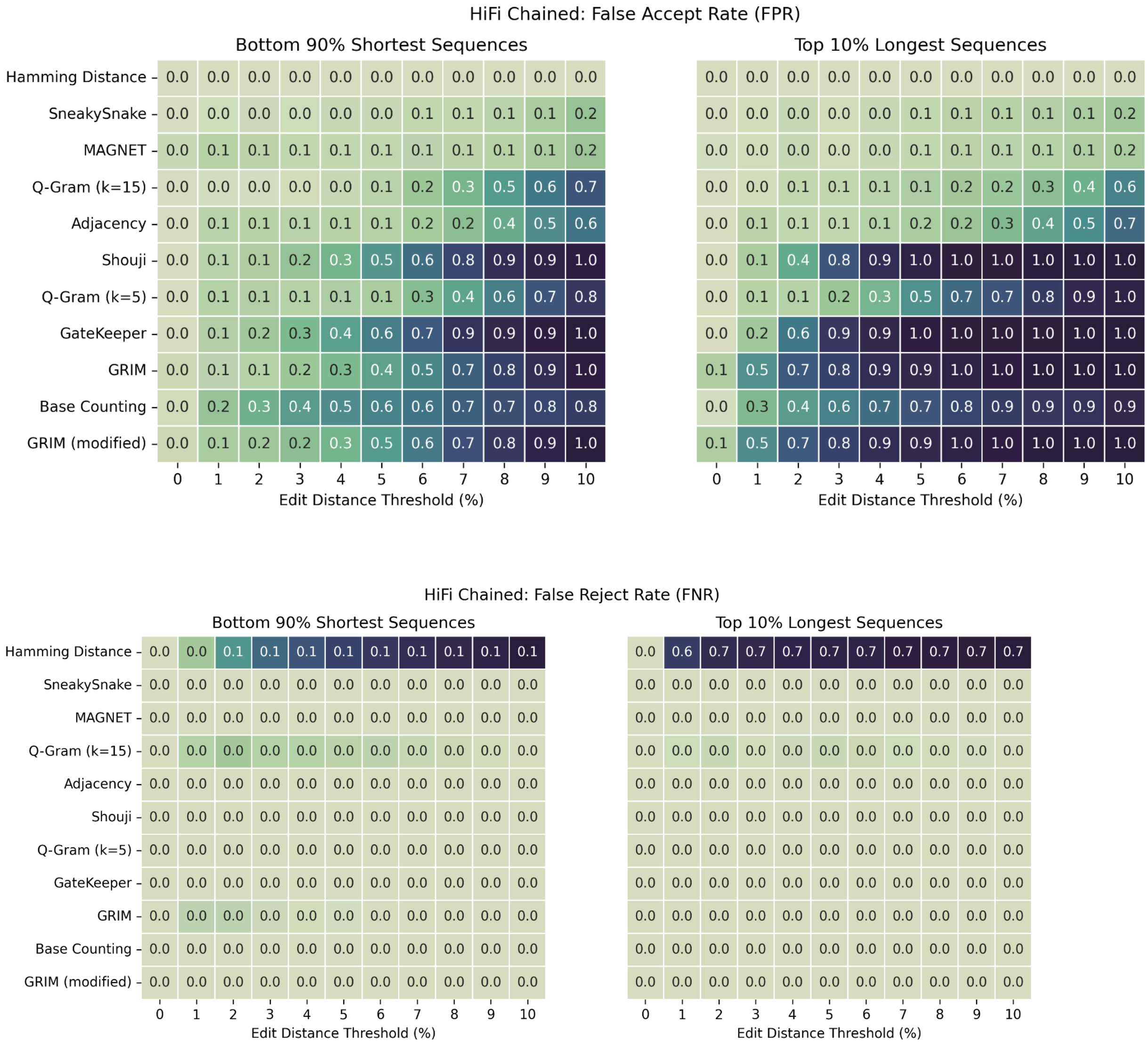

**Figure 7:** False Accept Rate (top) and False Reject Rate (bottom) for PacBio HiFi reads after chaining inside minimap2. 0.0 is the best, and 1.0 is the worst score. Note that boosting the False Accept Rate is easy by increasing the False Reject Rate – see Hamming Distance. FRR > 0 is undesirable as some correct pairs may be lost.

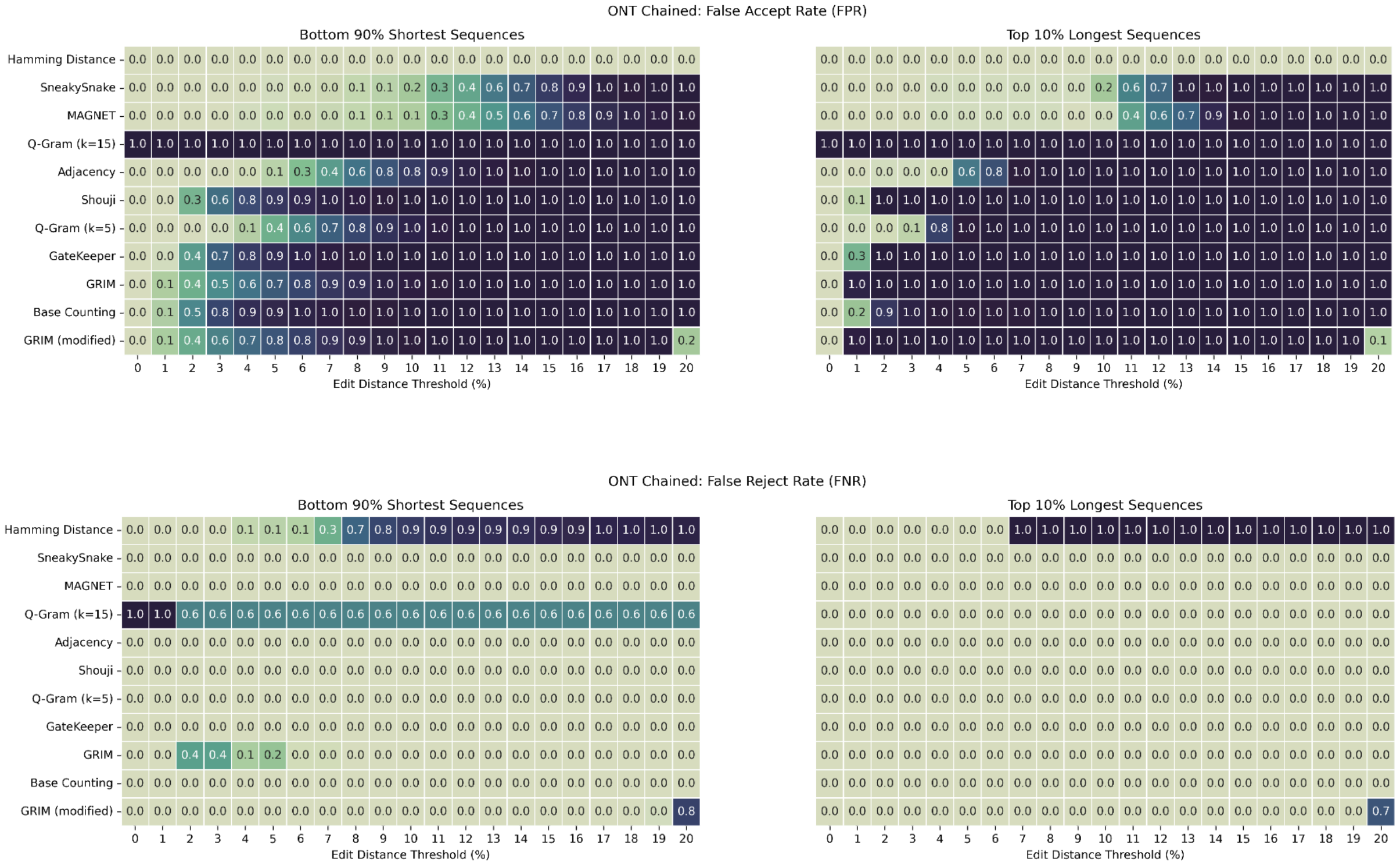

**Figure 8:** False Accept Rate (top) and False Reject Rate (bottom) for ONT reads after chaining inside minimap2. 0.0 is the best, and 1.0 is the worst score. Note that boosting the False Accept Rate is easy by increasing the False Reject Rate – see Hamming Distance. FRR > 0 is undesirable as some correct pairs may be lost.

## Execution Time

The goal of filtering is to reduce the overall computational burden. To evaluate performance, we analyze the execution time and throughput of the 11 filters and two aligners analyzed and see how it depends on the underlying data and the edit distance threshold. To maintain a fair comparison, we only tested CPU implementations of the algorithms provided. We reserve GPU, FPGA, and PIM benchmarks for future work.

### Longer Sequences Account for Most of the Execution Time

During our experiments, we observed that the filter throughput is highly sensitive to the length of the sequences. This effect was especially pronounced for filters whose runtime depends on the edit distance threshold. To investigate this effect, we split each of the six datasets: One containing the shortest 90% of the reads and the other the longest 10%. We break down the execution time between these splits in **Figure 9**. We make three key observations: 1) The quadratic complexity filters (Shouji, MAGNET, SneakySnake, GateKeeper) spend ~90% of their execution time on the longest 10% of sequences (81x longer on a per pair basis). 2) The linear complexity filters (q-gram, GRIM, and Base Counting), except

Hamming Distance, spend ~50% of their time on the longest sequences (9x longer on a per pair basis). 3) Adjacency and Hamming Distance spend time proportional to the share of the underlying data (~10% – indicated by the dotted line).

We conclude that most of the execution time comes from the longest sequences, even within the same dataset. Improving efficiency in this edge case or avoiding long sequences can dramatically increase performance. Additionally, as the sequence length is known beforehand, one could dynamically decide to use a more efficient $O(n)$ filter for the longest reads to achieve significant throughput improvements.

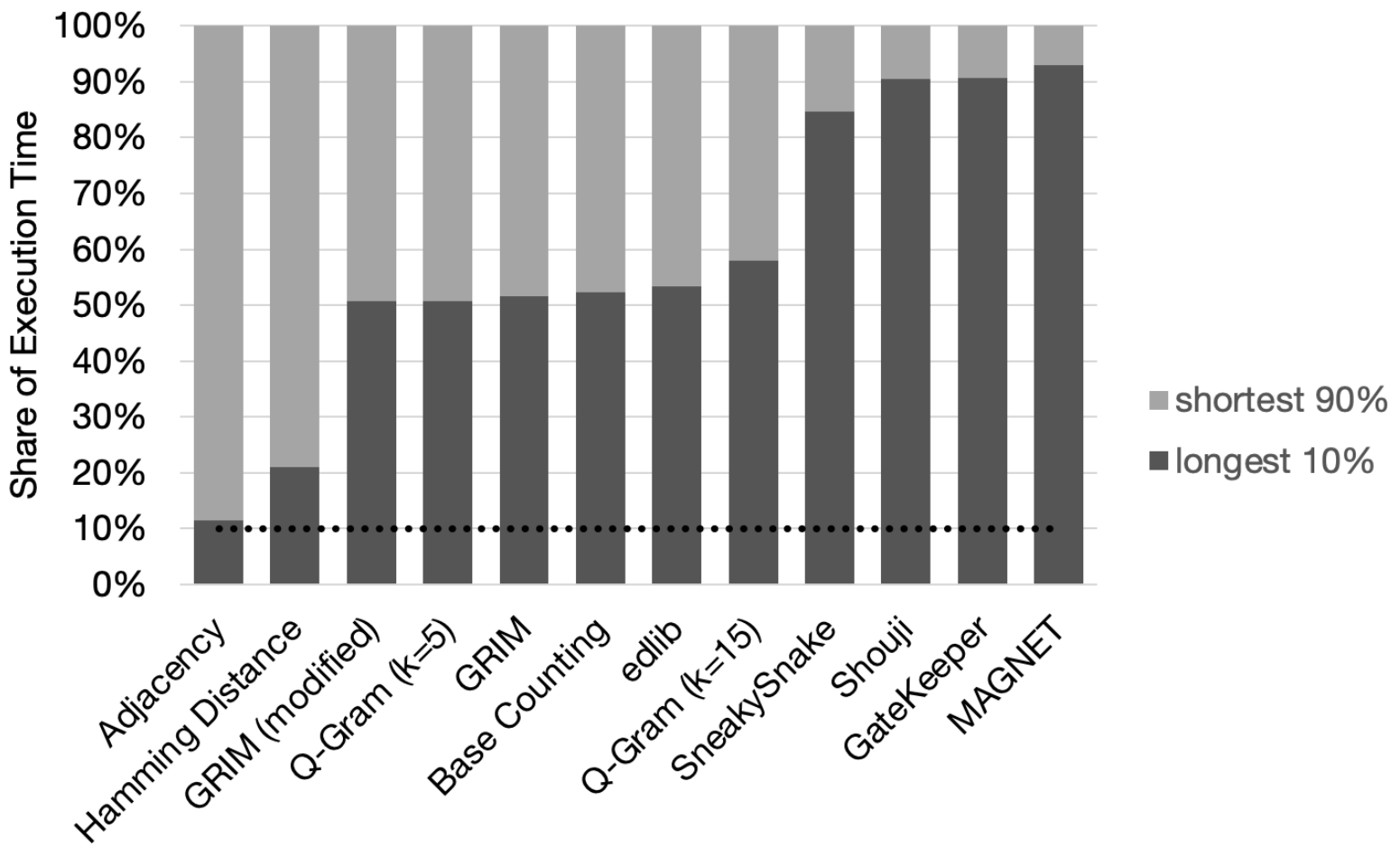

**Figure 9:** Execution time distribution between the longest 10% and shortest 90% for PacBio HiFi chained datasets. The dotted line at 10% indicates the share of the input data. Anything above that line indicates that computation for the longer sequences took more time.

**Filter Execution Time Varies Drastically Across Filters And Thresholds**

We analyze the total execution time across varying edit distance thresholds. We provide reduced plots to increase clarity, with full plots in the supplementary section. For each dataset, we provide the total execution time spent on the filter. We additionally provide log-log plots to observe scaling better: A horizontal line indicates the runtime does not scale with the edit distance, and a straight line with a greater than zero gradient implies polynomial scaling. For SneakySnake, we observe an even stronger scaling, but this is most likely due to the effects of SneakySnake's early termination.

We make the following four key observations from **Figure 10**: 1) All filters are significantly faster than the aligners (Edlib[125] shown) for Illumina and HiFi (except GateKeeper[104]). 2) Hamming Distance is the fastest filter across datasets. 3) We confirm quadratic scaling for SneakySnake, GateKeeper, Shouji, MAGNET and linear scaling for the others. 4) For ONT data, only a few filters (Hamming Distance shown) beat the performance of the aligners. We also note that many of the approaches are highly sensitive to implementation. All tested code was written in C/C++ without vector instructions. We remark that, e.g., PUNAS[110] has shown that order of magnitude improvements can be expected by correctly utilizing these instructions and other systems-level optimizations. We expect this to hold for other approaches, too.

In conclusion, most filters provide significant speedup over the aligners. Still, many suffer from quadratic time scaling with increasing edit distance thresholds and sequence length, making them unsuitable for long sequences. SneakySnake and Hamming Distance are very attractive candidates for Illumina. For HiFi and ONT, linear complexity filters dominate throughput, with Hamming Distance leading, followed by q-gram methods and Base Counting.

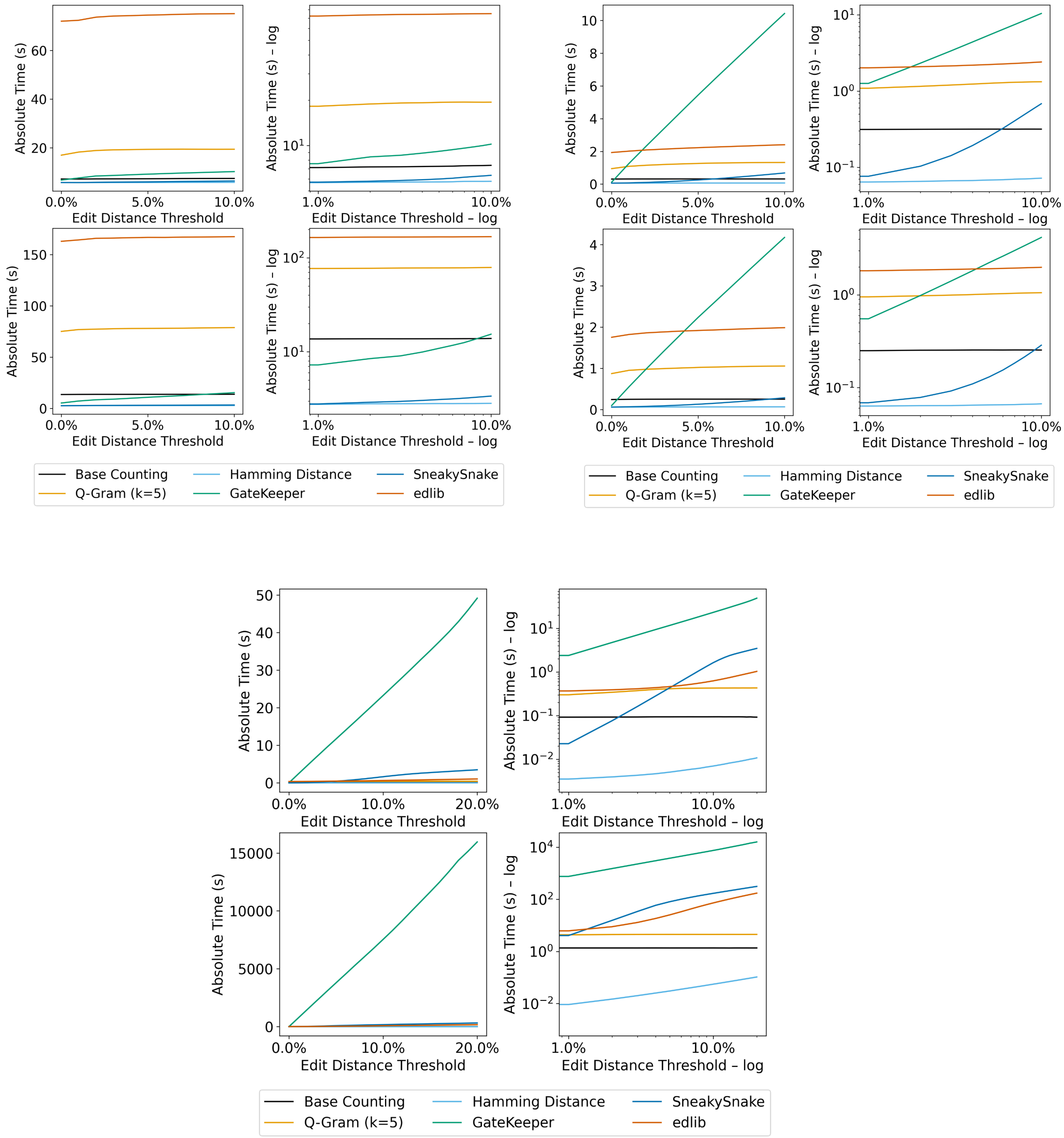

**Figure 10:** Runtime for Illumina (top left), HiFi (top right), and ONT (bottom left). Total filter time for 9M (Illumina), 90,000 (HiFi), and 4500 (ONT) sequence pairs. Linear axes on the left, log-log axes on the right. The top two charts in each cell are for chained data, and the bottom two are for mapped data.

## Workload Overlap

This is the first work to provide an overview of filter combinations. Many approaches, such as GASSST[117] and GenCache[108], use multiple pipelined filters but do not provide a standalone analysis. Combinations have the promise to improve overall accuracy and performance in two ways:

- **Accuracy**: Multiple filters may reduce the False Accept Rate by removing distinct sequence pairs. To analyze this, we provide a Venn diagram for the four highest-performing filters to see if they reject/accept the same set of pairs.
- **Performance**: Using a high-throughput filter first (such as Base Counting), we can reduce the workload passed to lower-throughput, higher-accuracy subsequent filters. We expect a significant reduction in the execution time without sacrificing accuracy based on observations from previous work[17].

We investigate the accuracy claim more closely and reserve the performance analysis for future work. If desired, a theoretical estimate of performance improvement can be calculated using the supplementary materials provided. In **Figure 11**, we provide a four-set Venn diagram illustrating the work overlap between filter combinations. We have chosen a mix of the most attractive individual filters for each sequencing technology based on throughput or accuracy. Using the data provided, they can be generated for an arbitrary filter combination using the supplementary materials. The sum of all percentages inside a circle indicates the share of rejections the filter contributed – higher is better. The percentage in an overlapping field indicates that any filter could have provided those rejections. A filter combination increases overall accuracy if there is a large share of "unique" contributions, i.e., a high percentage in the non-overlapping sections.

We make four key observations: 1) SneakySnake makes up ~99.7% of the rejection for Illumina, making multi-stage filters unattractive from an accuracy perspective. 2) For HiFi, q-gram methods and SneakySnake have a 79.3% overlap, meaning there is a moderate ~25% accuracy improvement to using a combination. 3) The simple Base Counting method is well-suited for HiFi, managing to filter 43.3% of the total. Using this as a pre-filter could generate significant speedup (future work). 4) For ONT, SneakySnake and q-gram (k=15) overlap strongly. Most other approaches are ineffective – not all are shown here.

We conclude that combinations of the current filters do not provide significant accuracy benefits for Illumina and ONT (Note: q-gram’s 17% is inflated due to false rejects). For HiFi, filter combinations are sensible, and we believe a Base Counting (applied first) and q-gram (k=15) combination might be fast and accurate even for long reads.

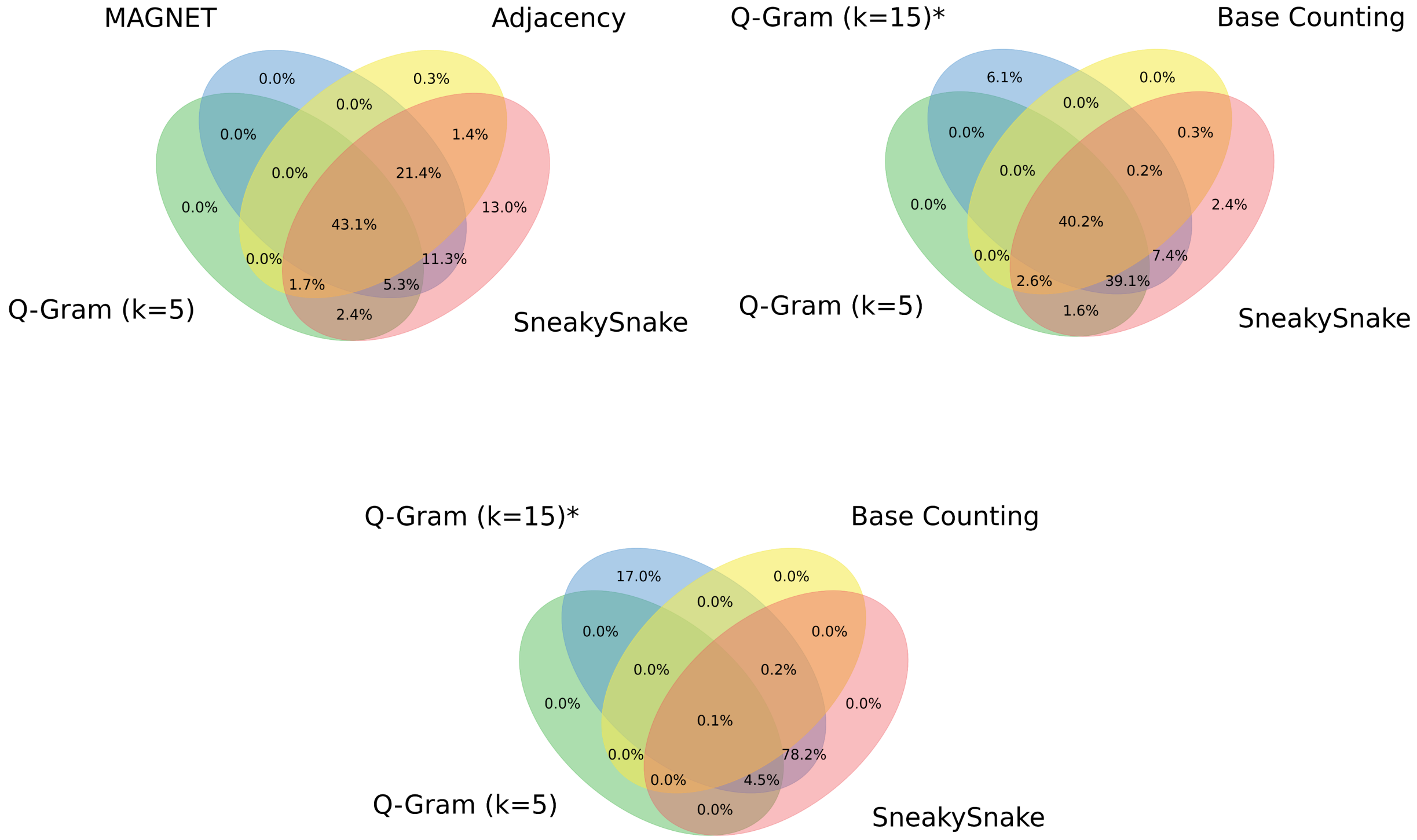

**Figure 11:** Venn diagrams showing the overlap between four methods representing different filtering methodologies. The number displayed is the *unique* number of pairs rejected by that filter or filter combination. Thus, high percentages are desirable, and low percentages indicate that a filter only removes a few additional sequences. To estimate how much work could be done by a filter or filter combination, sum the total percentages within the respective areas. Values were sampled at an edit distance threshold of 5% for Illumina (top left) & PacBio HiFi (top right) and 10% for ONT (bottom).

## Applications in Metagenomic Analysis

### Metagenomic Profiling

Microorganisms are ubiquitous in almost every natural setting, including soil, ocean water, and the human body. They play critical roles in the functioning of each of these systems. Traditional culture-based analysis of these microbes is confounded by many microorganisms that cannot be cultured in standard laboratory settings. Further, analysis of lab-cultured organisms fails to capture the complex community dynamics in real microbial ecosystems. Metagenomics, or the analysis of whole microbial genomes recovered directly from their host environment via high-throughput sequencing, is vital to understanding microbial communities and their functions. Taxonomic profiling, which predicts taxa's presence and relative abundance in a metagenomic sample, is a fundamental aspect of metagenomic analysis. It is computationally simpler and more effective in detecting low-abundance organisms than metagenomic assembly. Taxonomic profiles can be generated through read classification, where reads are assigned to specific taxa, or read binning, in which groups read into probable taxa or organism groups. However, due to the challenge of inferring taxa from short reads, these techniques have been demonstrated to result in lower accuracy in taxonomic profiling.

### Evaluating the benefits of pre-alignment filtering for read mapping

We evaluate the potential benefits of using pre-alignment filtering in metagenomic profiling. We introduce the previously studied pre-alignment filters to reduce the workload for the computationally expensive sequence alignment. We base our analysis on the state-of-the-art mapping-based metagenomic profiler Metalign. Metalign performs read-mapping using minimap2. We restrict our modifications of Metalign to the read-mapping stage. In the read mapping stage, we use each filter to estimate an edit distance value for all reads in the read set. We discard all reads with an approximate edit distance higher than 10% of the read length, as they are likely highly dissimilar to the subject species' genome. We pass all remaining reads, i.e., those with an estimated edit distance below the cutoff value, on to sequence alignment. The sequence alignment results are stored in a SAM file that Metalign uses to calculate the relative abundance of each species. We evaluate the benefits of pre-alignment filtering heuristics on real and simulated datasets. The Critical Assessment of Metagenome Interpretation (CAMI)[19,128] provides the most comprehensive and in-depth evaluation of metagenomic classifiers. The study examines metagenomic profiling methods based on diverse simulated metagenomic datasets. We produce the ground truth for our analysis by performing sequence alignment on the complete, unfiltered read set. Methodologically, this can be interpreted as a pre-alignment filter with its edit-distance cutoff threshold set to infinity, i.e., a filter that accepts all sequence pairs (from now on referred to as the *all accept* filter). Subsequently, we rely on the edit distance estimated by each filter to reject dissimilar sequences, thereby effectively filtering the read set. We evaluate the performance of all filters in terms of recall, precision, F1 score, Jaccard index, L1 norm error, and weighted UniFrac (full OPAL report in supplementary material). In total, we consider three datasets: a low-diversity (CAMI low) and a high-diversity (CAMI high) community from CAMI, each comprising 15 Gbp of sequence data and a real dataset from the TARA Ocean project (ERR1700889_1.fastq). The CAMI-high-complexity community includes numerous species not contained in the database: only 161 out of 243 unique species are present in our reference database.

We conduct metagenomic classification on these three datasets and perform an accuracy analysis using OPAL[128], the CAMI-affiliated evaluation software. As a representative example, we consider the results of the analysis as mentioned earlier based on the CAMI high complexity dataset. We run each filter with the same parameters and are exposed to the same workload. We remove all organisms assigned an abundance of less than 0.01% from all final profiles. For all read sets, we observe that all classification results are free of both false negatives and false positives, with each filter exhibiting an F1 score of one. We conclude that all methods are suitable as pre-alignment filters for metagenomic classification applications.

In the following evaluation, we assess the computational resources required by minimap2 when run with each filter (**Table 3**). For each filter, we provide the end-to-end runtime, peak RSS, the number of pairs accepted, and the total number of pairs processed by each filter. The number of processed pairs quantifies the workload for each filter. For comparability, we subject each filtering methodology to the same workload, meaning each filter processes the same number of pairs. The end-to-end runtime of the read mapping stage depends on the number of accepted sequences (the number of reads passed on to sequence alignment) and the share of execution time spent on the filter. The total runtime of Metalign differs only due to the different execution times in the read mapping stage, i.e., due to different filters. We thus restrict our runtime analysis to the read mapping stage only. The all-accept filter represents full sequence alignment. We make four key observations: (1) The number of invocations of each filtering heuristic is constant across filters and only depends on the read set. This confirms that all filters are exposed to the same workload, as intended. (2) The main memory footprint is practically independent of the choice of filtering method. (3) We evaluate pre-alignment filters in terms of accuracy by examining the number of accepted sequence pairs. Given no false negatives in the final taxonomic profile, we expect highly accurate filters to accept as few sequences as possible to achieve the greatest reduction in workload for later sequence alignment. We observe SneakySnake, MAGNET, and Edlib to reject the most sequences. (4) The pre-alignment filtering methods SneakySnake, Edlib, Hamming Distance, and Adjacency allow for the greatest speedup across all read sets. We achieve the best speedup of approximately 4x when benchmarked against full, unfiltered sequence alignment for CAMI high reads with SneakySnake and Hamming Distance.
We conclude that all examined pre-alignment filtering methodologies enable strong speedup of the state-of-the-art metagenomic profiler, Metalign, without compromising accuracy.

**Table 2:** A ranking of pre-alignment filters based on the total end-to-end runtime of minimap2, peak main memory usage, number of accepted reads (i.e., reads with an estimated edit distance below the defined threshold), and workload, with the workload being determined by counting the number of processed pairs for each filter. We repeat this analysis for each read set.

| CAMI low read set | | | | |
|---|---|---|---|---|
| **Filter** | **Runtime (mm.ss.ms)** | **Memory Footprint (B)** | **Accepted Pairs** | **Processed Pairs** |
| Adjacency | 25.55.50 | 7261108 | 1691181 | 8484773 |
| Edlib | 27.17.49 | 7261000 | 489965 | 8484773 |
| Base Counting | 28.32.40 | 7261100 | 4891290 | 8484773 |
| GRIM | 28.53.27 | 7261052 | 8484772 | 8484773 |
| GRIM (modified) | 29.53.98 | 7261136 | 8484772 | 8484773 |
| All Accept | 30.49.59 | 7260892 | 489957 | 8484773 |
| Shifted Hamming Distance | 31.14.50 | 7261260 | 6228291 | 8484773 |
| Q-Gram (k = 5) | 36.34.47 | 7261004 | 8484772 | 8484773 |
| SneakySnake | 38.06.54 | 7261044 | 768843 | 8484773 |
| Hamming Distance | 40.24.24 | 7260772 | 451443 | 8484773 |
| Shouji | 52.58.84 | 7261020 | 8366776 | 8484773 |
| Magnet | 58.58.57 | 7261100 | 735168 | 8484773 |
| CAMI high read set | | | | |
| **Filter** | **Runtime** | **Memory Footprint** | **Accepted Pairs** | **Processed Pairs** |
| Hamming Distance | 34.00.57 | 7260948 | 20927589 | 87266158 |
| SneakySnake | 39.14.35 | 7261012 | 25719195 | 87266158 |
| Adjacency | 52.09.47 | 7261144 | 28071919 | 87266158 |
| Base Counting | 1.00.41 | 7261072 | 59015999 | 87266158 |
| Edlib | 1.15.07 | 7260908 | 22530549 | 87266158 |
| Q-Gram (k = 5) | 1.29.38 | 7260868 | 87266157 | 87266158 |
| Shifted Hamming Distance | 1.33.39 | 7260900 | 68199860 | 87266158 |
| All Accept | 2.00.57 | 7260864 | 22530538 | 87266158 |
| GRIM | 2.00.58 | 7260924 | 87266157 | 87266158 |
| GRIM (modified) | 2.00.59 | 7260712 | 87266157 | 87266158 |
| Shouji | 2.04.10 | 7261060 | 83499414 | 87266158 |
| Magnet | 2.08.57 | 7260928 | 25596661 | 87266158 |
| TARA Ocean read set | | | | |
| **Filter** | **Runtime** | **Memory Footprint** | **Accepted Pairs** | **Processed Pairs** |
| Edlib | 28.54.70 | 7261008 | 1246472 | 1794985 |
| Shouji | 28.54.52 | 7261132 | 1547258 | 1794985 |
| SneakySnake | 29.13.81 | 7261096 | 1288836 | 1794985 |
| Hamming Distance | 29.54.48 | 7261036 | 1226234 | 1794985 |
| Q-Gram (k = 5) | 30.54.93 | 7261008 | 1555669 | 1794985 |
| Shifted Hamming Distance | 31.55.07 | 7261132 | 1501273 | 1794985 |
| GRIM | 32.59.55 | 7260920 | 1555669 | 1794985 |
| GRIM (modified) | 34.10.21 | 7260816 | 1555669 | 1794985 |
| All Accept | 38.14.08 | 7261064 | 1245332 | 1794985 |
| Base Counting | 39.03.07 | 7261024 | 1378134 | 1794985 |
| Adjacency | 39.17.82 | 7261036 | 1398847 | 1794985 |
| Magnet | 49.01.53 | 7261076 | 1284956 | 1794985 |

## Discussion and Future Work

Filtering heuristics are vital for handling large genomic datasets and accelerating their applications, making them more accessible. Our review focuses on the interplay between technological and algorithm development. It guides the choice of the appropriate algorithm and identifies new algorithmic research directions in response to the advancement of long-read technologies and novel sequencing protocols. Given the importance and prominence of genomic sequence comparison across domains and applications, we anticipate that our study will be a valuable resource for many academic and industrial research groups performing broad biomedical research.

### Directions for improving genomic sequence comparison

There is a great need for new filters to address our uncovered shortcomings. There are no satisfactory filters for ONT and HiFi. Five directions hold promise: 1) Making a smart tradeoff between having a slightly above zero false reject rate. This should allow designers to decrease FAR significantly. The benchmarking suite allows designers to measure this tradeoff and find a great balance. 2) Use filter combinations for HiFi data to improve overall accuracy and performance. 3) Develop entirely new algorithms to tackle these new data types. 4) Accelerate existing low FAR but slow filters using FGPAs, GPUs, or Processing-in-Memory. 5) Dynamically adapting which filtering technique to use based on the edit distance threshold, sequence length, and technology type.

## Availability of data and materials

We exclusively used publicly available datasets in this paper. The CAMI results from our metagenomic analysis are based on the CAMI challenge datasets. Information on CAMI challenge data is available on the official CAMI Website or can be downloaded from the GigaDB website. The prokaryote-isolated Tara Oceans reads used in our study on metagenomic applications are available on EBI: https://www.ebi.ac.uk/ena/data/view/PRJEB1787. The run accessions are ERR598952 and ERR598957.
SequenceLab source code is available on GitHub: https://github.com/CMU-SAFARI/SequenceLab.

## Competing interests

The authors declare that they have no competing interests.

## Key Points

- SequenceLab is a comprehensive benchmark suite dedicated to analyzing the most important algorithms used to compare genomic sequences.

- We also evaluate the applicability of heuristic methods as pre-alignment filters in read mapping for taxonomic profiling of metagenomic samples.

- We investigate if combining heuristic methods enhances performance over single-method use.

- Our study assesses how different genomic data characteristics impact filtering efficacy. We establish our evaluation based on Illumina, PacBio HiFi, and ONT datasets, realizing a comprehensive evaluation with a broad range of real-world applications.

## Author Biographies

**Maximilian-David Rumpf** is a member of the Department of Computer Science at ETH Zürich. His research focuses on computer architecture, artificial intelligence, and bioinformatics. He is also the founder of SID.ai – a San Francisco-based technology startup backed by YCombinator.

**Mohammed Alser** is a senior researcher and lecturer of bioinformatics and computer architecture at D-INFK and D-ITET, ETH Zürich. His primary research incorporates several aspects of bioinformatics, metagenomics, computational genomics, and computer architecture. He is particularly interested in rethinking the complete compute stack, from how we handle input data and algorithms to the underlying hardware architecture for efficiently analyzing genomic data on existing and unconventional (e.g., non-von Neumann) computing technologies.

**Arvid E. Gollwitzer** is an entrepreneur and graduate student at ETH Zurich. His current research interests are in bioinformatics, clinical genomics, computer architecture, and the early detection of cancer. Throughout his graduate studies in Electrical Engineering and Information Technology at ETH Zurich, he specialized in clinical metagenomics and early-stage cancer detection. Arvid also co-founded NextGuide Medical, where he develops AI assistance technology for visually impaired individuals. His professional experience, teaching, and graduate studies include roles at CERN and the SAFARI Research Group of ETH and CMU. Arvid has published several works on bioinformatics, metagenomics, and clinical genomics. He also holds an excellence scholarship from the Swiss Study Foundation, an organization dedicated to promoting the top 3% of students based on their exceptional academic impact, creativity, and character.

**Joël Lindegger** is a PhD student at the ETH Zurich SAFARI research group. His research interests are accelerating the bioinformatics pipeline, current and future computer architectures, and algorithms and data structures.

**Nour Almadhoun Alserr** is a senior researcher at ETH Zurich. Her research focuses on data privacy and bioinformatics. She obtained her PhD in Computer Engineering at Bilkent University.

**Can Firtina** is a Ph.D. student in the SAFARI Research Group at ETH Zurich, advised by Prof. Onur Mutlu. He obtained his B.Sc. and M.Sc. degrees in Computer Engineering from Bilkent University.
His current research interests broadly span bioinformatics and computer architecture topics, including accurately and quickly identifying sequence similarities, real-time genome analysis, hardware/software co-design for accelerating bioinformatics workloads, correcting sequencing errors, and developing computational tools for genome editing.

**Serghei Mangul** is an assistant professor in the Titus Family Department of Clinical Pharmacy at USC Mann. He completed a postdoctoral fellowship at UCLA's Institute for Quantitative and Computational Biosciences. Before that, he was a visiting scholar at Harvard Medical School after he earned his PhD in bioinformatics at Georgia State University.

**Onur Mutlu** is a Professor of Computer Science at ETH Zurich. He is also a Visiting Professor at Stanford University and an Adjunct Professor at Carnegie Mellon University. He is especially interested in interactions across domains and between applications, system software, compilers, and microarchitecture, with a major current focus on memory and storage systems, bioinformatics, and biologically inspired computation paradigms. He holds a PhD from the University of Texas at Austin and BS degrees in Computer Engineering and Psychology from the University of Michigan, Ann Arbor. His professional background includes roles at Microsoft Research, Intel, and Google. He is recognized as an ACM Fellow and IEEE Fellow for his substantial contributions to computer architecture, particularly in memory systems. Onur Mutlu's academic influence is highlighted by numerous awards, including the 2023 Google Open Source Peer Bonus Award and the 2020 IEEE Computer Society Edward J. McCluskey Technical Achievement Award. He is also lauded for his educational contributions, with his lectures available on YouTube and numerous resources provided by his research group, SAFARI.